\documentstyle[psfig,aps,prl,twocolumn]{revtex}

\begin{document}
\draft \author{Immanuel Bloch, Michael K\"ohl, Markus Greiner,
Theodor W.~H\"ansch, and Tilman Esslinger}
\address{Sektion Physik,
Ludwig-Maximilians-Universit\"at, Schellingstr.\ 4/III, D-80799
Munich, Germany and\\
Max-Planck-Institut f\"ur Quantenoptik, D-85748 Garching, Germany}
\title{Optics with an Atom Laser Beam}
\maketitle
\begin{abstract}

\noindent

We report on the atom optical manipulation of an atom laser beam.
Reflection, focusing and its storage in a resonator are
demonstrated. Precise and versatile mechanical control over an
atom laser beam propagating in an inhomogeneous magnetic field is
achieved by optically inducing spin-flips between atomic ground
states with different magnetic moment. The magnetic force acting
on the atoms can thereby be effectively switched on and off. The
surface of the atom optical element is determined by the resonance
condition for the spin-flip in the inhomogeneous magnetic field.
More than $98\,\%$ of the incident atom laser beam is reflected
specularly.

\end{abstract}

\pacs{03.75.Fi, 03.75.Be, 07.77.Gx, 32.80.-t}

\noindent The realization of atom lasers\,\cite{atomlasers} opens
up new intriguing perspectives in coherent atom optics. These
novel atom sources are based on Bose-Einstein
condensates\cite{bec} from which a coherent matter wave beam is
extracted. The unique properties of atom lasers will make it
possible to enter an experimental regime in atom optics that is
not accessible for thermal atom sources. In light optics the
availability of coherent sources has substantially increased the
range of photonic applications. Similarly, it is expected, that
coherent matter wave sources will have a profound impact on
applications such as atom
interferometry\,\cite{atominterferometry}, atom
holography\,\cite{atomholography} or the manipulation of atomic
beams on a nanometer scale.

To fulfill these expectations it is crucial to invent atom optical
elements that are adapted to the demands of the new highly
coherent atom sources. For thermal atom sources a variety of atom
optical elements have been investigated\,\cite{Adams94}, which
redirect, split or shape atomic beams using position- or
time-dependent potentials. To preserve the coherence properties of
atom lasers a very high "surface quality" of atom optical elements
is required. The de Broglie wavelength of an atom laser beam can
be well below ten nanometers. The effective surface roughness of
atom optical elements should therefore be even smaller.

The most simple approach to realize a coherent matter wave source
is to suddenly release a Bose-Einstein condensate from the
magnetic trap. The mechanical manipulation of released condensates
has been demonstrated using optical standing wave
fields\,\cite{Phillips99,Stenger99} suitably shaped off-resonant
laser fields\,\cite{Ertmer99} and pulsed magnetic
fields\,\cite{Boshier99}. Due to the sudden switch-off of the
confining potential the energy of the repulsive interaction
between the atoms is transformed into kinetic energy. The velocity
distribution of the released atoms is therefore much broader than
the Heisenberg limit associated with the spatial size of the
trapped condensate. The velocity spread is drastically reduced for
atom lasers employing continuous output coupling, where a Fourier
limited output can be approached\,\cite{Band99} and interaction
effects are minimized. The resulting monoenergetic atom laser beam
is not susceptible to dispersive effects in the manipulation of
coherent matter waves\,\cite{Ertmer99}, which are unwanted in most
atom optical applications.

In this Letter we report on the atom optical manipulation of an
atom laser beam. We have realized a versatile atom optical element
and demonstrate reflection, splitting and focusing of the atom
laser beam, as well as its storage in a resonator. This is
accomplished by optically inducing spin-flips between atomic
ground states of different magnetic moment, thereby switching the
force on and off that the atom laser beam experiences in an
inhomogeneous magnetic field. By a suitable choice of the
frequencies and polarizations of the Raman laser fields, the
coupling between two specific ground states can be induced. This
avoids restrictions in the coupling strengths and state
selectivity that would be encountered when using rf or microwave
fields.

The atom laser output is extracted from a $^{87}$Rb Bose-Einstein
condensate using continuous output coupling\,\cite{Bloch99}. A
weak and monochromatic radio-frequency field transfers the
magnetically trapped atoms, which are condensed in the ${|F=1,
m_F=-1 \rangle}$ state, into the untrapped $|F=1, m_F=0 \rangle$
state ($F$: total angular momentum, $m_F$: magnetic quantum
number). Gravity accelerates the untrapped atoms downwards and a
well collimated atom laser beam is formed, while the magnetic trap
is still on. After ballistically propagating over a few hundred
micrometers the atoms enter the spin-flip region, where two
focused laser beams induce a two photon hyperfine Raman transition
and transfer a variable fraction of atoms into the state $|F=2,
m_F=1 \rangle$ with the magnetic moment $\frac{1}{2}\,\mu_B$
($\mu_B$: Bohr magneton).

\begin{figure}
\centerline{\psfig{file=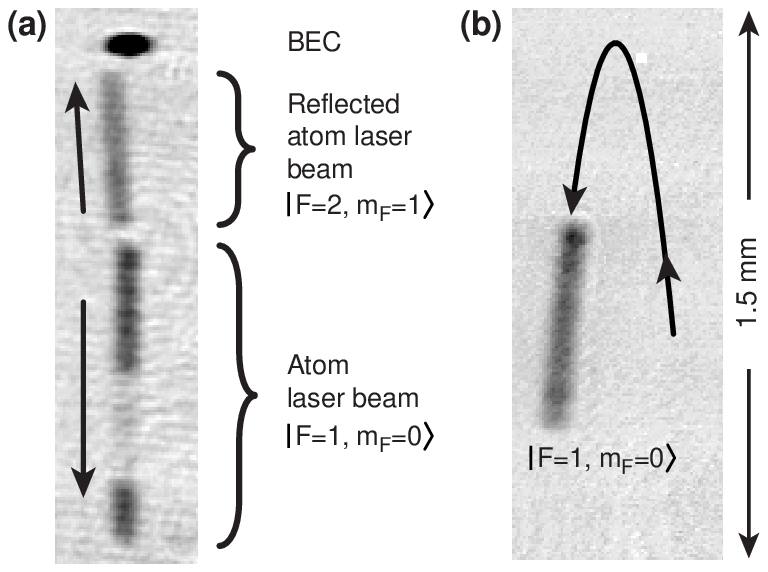,angle=0,width=0.4\textwidth}}
\caption {Reflected atom laser beams for a single {\bf (a)} and
double pass {\bf (b)} through the Raman lasers. {\bf (a)} The
spin-flip mirror was switched on during a period of 2\,ms. An
adjustable fraction of the atom laser beam is spin-flipped into
the $|F=2, m_F=1\rangle$ state, reflected and moves upwards, as
indicated by the arrow. The unaffected parts of the atom laser
beam propagate downwards. {\bf (b)} The spin-flip mirror was
switched on for a sufficiently long time, so that all atoms that
were spin-flipped into the $|F=2,m_F=1\rangle$ state on their way
downwards were spin-flipped again back into the
$|F=1,m_F=0\rangle$ state during their propagation upwards.The
angle under which the beams are reflected is caused by a weak
horizontal component of the magnetic field gradient at the
position of the spin-flip mirror.} \label{reflection}
\end{figure}

\begin{figure}
\centerline{\psfig{file=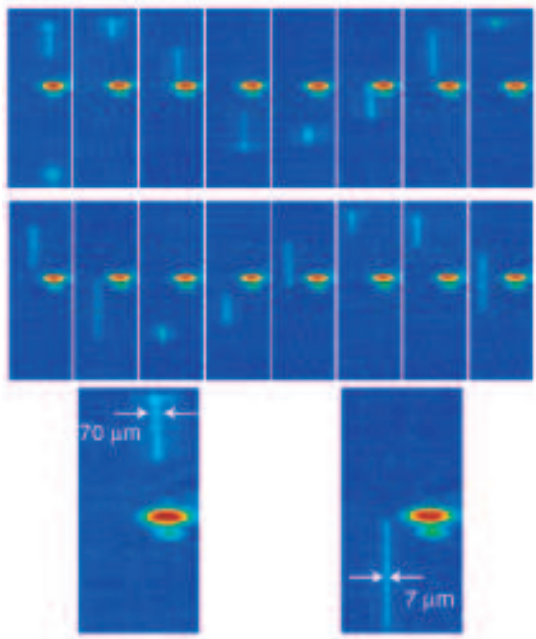,angle=0,width=0.4\textwidth}}
\caption {Atom laser beam stored in a resonator formed by the
magnetic trapping potential. The elliptical cloud in the centre of
each absorption image is the condensate. The stripe of varying
width and length is the atom laser beam, which oscillates in the
trap. Part of the atom laser beam which has not been recaptured
can be seen in the first image of the upper row. Each image has a
size of 2\,mm$\times$0.7\,mm. The two images in the lowest row
show enlargements of the first image in the first row and the
second image in the second row. The propagation time was increased
by 2\,ms for each image.} \label{resonator}
\end{figure}

The state $|F=2, m_F=1 \rangle$ is low-field seeking and the atoms
experience the potential of the magnetic trap. The gradient of our
trap in the vertical direction exceeds the gravitational force by
one order of magnitude. Therefore the atoms are slowed down until
they reverse their direction of motion. Traveling upwards the
atoms pass through the laser field for a second time where they
are spin-flipped into the original state $|F=1, m_F=0 \rangle$ and
continue their motion on a ballistic trajectory (see
Figure\,\ref{reflection}(b)). When the Raman lasers are switched
off before the atoms cross the interaction region a second time,
the atoms remain in the magnetically trapped state, see
Figure\,\ref{reflection}(a) and Figure\,\ref{resonator}.

\begin{figure}
\centerline{\psfig{file=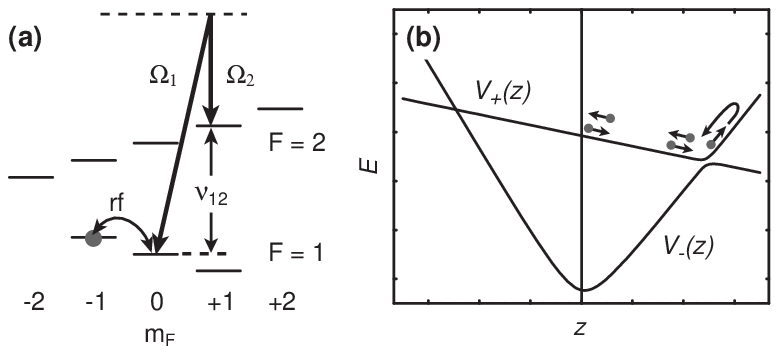,angle=0,width=0.5\textwidth}}
\caption { {\bf(a)} Level scheme of the $5S_{1/2}$ hyperfine
ground state of $^{87}$Rb in a magnetic field (not to scale). The
condensate is produced in the $|F=1, m_F=-1\rangle$ state and the
atom laser is generated by coupling the condensate to the $|F=1,
m_F=0\rangle$ state using an rf transition. The Raman lasers drive
the two-photon transition between the $|F=1, m_F=0\rangle$ and
$|F=2, m_F=1\rangle$ state and are off resonance with the single
photon excitation to the $5P_{1/2}$ state.{\bf(b)} Adiabatic
potentials $V_+(z)$ and $V_-(z)$ in the presence of the Raman
lasers.}\label{Ramantransition}
\end{figure}

The frequencies of the two laser fields that induce the Raman
transitions are adjusted to drive the single photon 5S$_{1/2}$ to
5P$_{1/2}$ transition off resonance (see
Figure\,\ref{Ramantransition}(a)). The polarizations of the laser
beams are chosen such that the lower frequency field drives
$\pi$-transitions and the higher frequency field drives
$\sigma$-transitions. Here the quantization axis is chosen
parallel to the local magnetic field vector, which is oriented
vertically in the spin-flip region. For both laser fields the
single photon detuning $\Delta$ is large compared to the single
photon Rabi frequencies $\Omega_{1}$ and $\Omega_{2}$. Spontaneous
scattering of photons is therefore suppressed and the coupling
between the states $|F=1, m_F=0 \rangle$ and $|F=2, m_F=1 \rangle$
can be described by an effective two level system, with a coupling
strength of $ \overline{\Omega}_0 = - \frac{\Omega_1
\Omega_{2}^{*}}{2 \Delta}$ \cite{Shore}.

The resonance condition for the spin-flip transition is fulfilled
if the frequency difference between the two Raman lasers is equal
to the hyperfine plus Zeeman splitting of the two atomic states.
Co-propagating laser beams are used in the experiment so that the
Raman transition is not sensitive to the Doppler effect. The
resonance condition mentioned above is fulfilled on a shell of
constant magnetic field. In other words, the effective surface of
the mirror in the inhomogeneous trapping potential is determined
by the frequency difference between the two Raman laser beams,
which can be controlled with high accuracy. The spin-flip
therefore occurs in a region which is much better localized than
the waist of the Raman laser beams. The distance between the
spin-flip region and the coils, which produce the dc-magnetic
field, is a few centimeters. Therefore static corrugations of the
mirror surface on a smaller spatial scale, e.g. in the nanometer
range, are not expected. To minimize fluctuations in the magnetic
trapping field a highly stable current supply ($\Delta
I/I<10^{-4}$) is used and the trapping region is placed in a
magnetic shield enclosure.

In the experiment Bose-Einstein condensates of typically $7 \times
10^5 $ $^{87}$Rb atoms are produced in a quadrupole and Ioffe
configuration (QUIC) trap\,\cite{Esslinger98} by evaporative
cooling. The laser light used to drive the two photon Raman
transition is generated by two extended cavity diode
lasers\,\cite{Ricci}. A phase-locked-loop\,\cite{Phaselock} is
employed to stabilize the frequency difference between the two
lasers to a frequency reference, which is tunable at around
6.8\,GHz corresponding to the hyperfine splitting of the $^{87}$Rb
ground state. Each of the phase-locked lasers is amplified by
injection locking another laser diode. The amplified laser beams
pass through acousto-optic modulators (AOM) used for switching
them on and off. Then the two beams are overlapped with orthogonal
polarizations and fed into a single mode optical fiber. The fiber
filters the spatial mode and ensures that the laser beams are
exactly co-propagating. After the fiber the laser beams are
directed into the vacuum chamber and propagate along the symmetry
axis (y-direction) of the elongated magnetic trapping potential,
in which the cigar shaped condensate is stored. The focus of the
overlapping laser beams is positioned $400\,\mu$m below the
condensate at $z_0=-400$\,$\mu$m. The beam waists are
$w_z=27$\,$\mu$m in the vertical and $w_x=500\,\mu$m in the
horizontal direction.

\begin{figure}
\centerline{\psfig{file=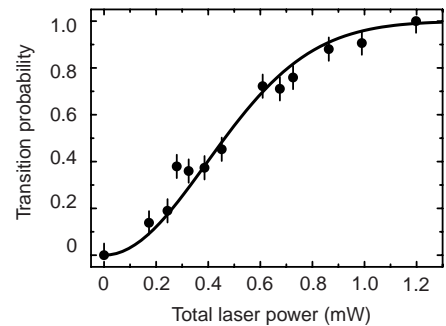,angle=0,width=0.35\textwidth}}
\caption {Reflectivity of the spin-flip mirror. The curve shows
the probability for an adiabatic transition as a function of the
total intensity of the Raman laser beams. The full line is a fit
to the data using the Landau-Zener model, as explained in the
text. The scaling between the squared coupling strength
$\overline{\Omega}_0^2$ and the laser intensity is taken as a free
parameter.} \label{reflectivity}
\end{figure}

We experimentally determined the reflectivity of the spin-flip
mirror, with the Raman lasers detuned by $70$\,GHz to the red of
the $D_1$ line of $^{87}$Rb and a magnetic field gradient of
200\,G/cm\,\cite{gradient}. An atom laser beam was extracted from
the condensate for $4$\,ms. The Raman lasers were switched on
$8.5$\,ms after the beginning of the output coupling for a
duration of $2$\,ms. During this period of time a fraction of the
atom laser beam was spin-flipped into the $|F=2, m_F=1 \rangle$
state and reflected. $5$\,ms later the magnetic trap was switched
off and an absorption image was taken (see
Figure\,\ref{reflection}(a)). The reflectivity of the mirror, i.e.
the probability for spin-flipping the atoms, was derived from the
absorption images. The reduced optical density in that part of the
atom laser beam which passed through the spin-flip region when the
Raman lasers were switched on was compared to the optical density
of the unperturbed atom laser beam. In addition, the optical
density of the reflected, i.e. spin-flipped, part of the atom
laser beam was measured. No loss of atoms was found in the
reflection process. Figure\,\ref{reflectivity} shows the mirror
reflectivity vs. the power of the Raman laser beams. For laser
powers of $1.2$\,mW a peak reflectivity in excess of $98\%$ was
found.

The measured reflectivity can be described by adiabatic
transitions according to a Landau-Zener model. The adiabatic
potential curves for the atoms in a one-dimensional case are given
by

\begin{equation}
V_{\pm}(z) = \frac{1}{2} \left( V_{1}(z) + V_{0}(z) \pm \sqrt{4
\hbar^2 \overline{\Omega}^2(z) + \Delta_{12}^2(z)} \right),
\end{equation}
where $ V_{1}(z)= m g z + h \nu_{hf} + 1/2 \mu_B
\sqrt{B_0^2+(B^{\prime} z)^2} $ and $ V_{0}(z)= m g z$ are the
potentials for atoms in the ${|F=2, m_F=1 \rangle}$ and ${|F=1,
m_F=0 \rangle}$ state, respectively (m: mass of $^{87}$Rb, $g$:
gravitational acceleration, $h$: Planck's constant). $B_0$ and
$B^{\prime}$ denote bias field and radial gradient of the magnetic
trap, which has a Ioffe-type magnetic field geometry (see
Figure\,\ref{Ramantransition}(b)).

The frequency difference $\nu_{12}$ of the Raman lasers is detuned
from the energy splitting of the trapped and untrapped atomic
states by an amount $\Delta_{12}(z)= V_{1}(z) -
V_{0}(z)-\Delta_{AC}(z) -h\nu_{12}$, where $\Delta_{AC}$ is the
residual difference in light shift induced by the Raman lasers,
that the atomic states experience. The spatial dependence of the
Raman coupling is determined by the gaussian laser focus
$\overline{\Omega}(z)=\overline{\Omega}_0 e^{-2(z-z_0)^2/w_z^2}$.

The atoms pass through the interaction region with a velocity $v$
which gives rise to non-adiabatic behaviour. The probability for
non-adiabatic passage is given by

\begin{equation}
p_{n. a.} = e^{-2 \pi \Gamma},
\end{equation}
with

\begin{equation}
\Gamma = \hbar \frac{\overline{\Omega}_0^2}{1/2 \mu_B B^{\prime}
v}.
\end{equation}
The efficiency of the adiabatic transition can thus be controlled
by the atomic velocity $v$ and $\overline{\Omega}$, which is
proportional to the laser intensity. For the narrow longitudinal
velocity spread of the atom laser beam the dependence
 on velocity
of the Landau-Zener transition is negligible. In a numerical
simulation we have verified that the finite interaction time of
the atoms with the laser beam and its gaussian shape results only
in a slightly modified effective coupling strength and preserves
the general form of the transition probability for our
experimental parameters.

The optical access in the experiment allowed us to demonstrate
reflection of the atom laser beam for various dropping heights of
up to 0.8\,mm. However, it is instructive to examine how the laser
power required for adiabatic transitions scales with the dropping
distance. From the criterion for adiabaticity
$\frac{\overline{\Omega}_0 \tau}{2 \Gamma}>
1$\,\cite{LandauZenercorrections}, where $\tau = w_z/v$, we can
estimate that the required Raman laser power $P= \sqrt{P_1 \times
P_2}$ scales as $P\propto \sqrt{z_0}$, where $z_0$ is the dropping
distance in the gravitational potential. Since for dropping
heights of $0.5$\,mm only $1$\,mW of laser power is required,
applying this scheme to much larger dropping heights is realistic.
This is in contrast to atom optical mirrors using optically
induced dipole potentials, where several watts of laser power are
needed for reflection of atoms from the same dropping
distance\,\cite{Ertmer99}. Furthermore, in the latter case the
required laser power scales as $P\propto z_0$.

We have experimentally determined the specularity of the
reflection process by comparing the transverse width of the atom
laser beam before and after the reflection process. We have found
that after a propagation time of 18\,ms the width decreases by
$25\,\pm 5\%$, which is in accordance with the weak curvature of
the mirror in axial direction. The transverse velocity spread of
the atoms due to the reflection is well below 200 $\mu$m/s, which
is the resolution limit in our experimental geometry. We can thus
conclude that the reflection is completely specular within this
limit. In comparison, evanescent wave mirrors or magnetic surface
mirrors severely suffer from substantial diffuse
reflection\,\cite{Aminoff93,Hinds99}.

To demonstrate the versatility of the atom optical spin-flip
element we have demonstrated the storage of the atom laser beam in
the resonator formed by the magnetic trapping potential. This was
achieved by choosing a timing sequence for the Raman lasers such
that the atoms in the atom laser beam were spin-flipped only once.
We could then observe the storage of the atom laser beam  in the
resonator formed by the magnetic trapping potential (Figure\,
\ref{resonator}). In this configuration we could monitor the atom
laser beam for more than $35$ oscillation periods. Due to the
axial curvature of the magnetic trapping potential  the atom laser
beam was focused. The initial width of the atom laser beam is
$70$\,$\mu$m, which is determined by the axial size of the BEC
ground state in the magnetic trap. After a propagation period of
$18$\,ms a width of the atom laser beam of only $7$\,$\mu$m,
limited by the resolution of our imaging system, is measured.

Focusing of the atom laser beam is a central step towards an atom
laser microscope\,\cite{Balykin}. To achieve a highly
monoenergetic atom laser beam of short de Broglie wavelength our
novel atom optical element could be used to accelerate the beam.
It should then be possible to focus the beam to spot sizes much
smaller than achievable in confocal light microscopy and
comparable to high energy electron microscopy. An atomic fountain
geometry as required for atomic clocks and atom
interferometers\,\cite{Kasevich} could be realized by shaping
magnetic field gradients and cascading atom optical spin-flip
elements.

In conclusion, we have demonstrated and quantitatively studied a
versatile atom optical element which allows manipulation of an
atom laser beam with unprecedented precision in a free space
environment. The experiments are crucial for future studies of
fundamental properties of the atom laser and its applications.

\begin{references}
\bibitem{gradient} When other values for the magnetic field gradient were chosen no
 fundamentally different behaviour was found.
\end{references}
\end{document}